\documentclass[A4paper, 12pt]{article}

\usepackage{hyperref}
\usepackage[utf8]{inputenc}
\usepackage[OT2,T1]{fontenc}
\usepackage[english]{babel}
\usepackage[osf]{mathpazo}
\usepackage{graphicx}
\usepackage{amsmath}
\usepackage{upgreek}

\usepackage{color}
\definecolor{FOB}{rgb}{0.,0.,0.7}
\definecolor{FOR}{rgb}{0.65,0.,0.}
\definecolor{FOfond}{rgb}{0.98,0.98,0.91}

\pagecolor{FOfond}

\usepackage{sectsty}
\allsectionsfont{\color{FOR}}

\usepackage{amssymb}
\usepackage{authblk}

\def\eqref#1{(\ref{#1})}
\def\Description#1{\relax}

\def\dd{{\rm d}}

\def\K{{\bf K}}
\def\N{{\bf N}}

\def\C{{\bf C}}
\def\R{{\bf R}}

\def\Z{{\bf Z}}

\def\bull{\vrule height .9ex width .8ex depth -.1ex }

\def\acosh{{\rm acosh}}
\def\GCD{{\rm GCD}}

\newcounter{thenum}
\def\texttheo{\relax}
\newenvironment{theorem}{\medbreak\refstepcounter{thenum}
\noindent\textsc{Theorem} %
\thethenum. \texttheo ---  \it  }{\rm }
\newenvironment{algorithm}{\medbreak\refstepcounter{thenum}
\noindent\textsc{Algorithm} %
\thethenum. \texttheo ---  \rm  }{\rm }
\newenvironment{e-proposition}{\medbreak\refstepcounter{thenum}
\noindent\textsc{Proposition} \thethenum. ---  \it  }{\rm }

\newenvironment{e-definition}{\medbreak\refstepcounter{thenum}
\noindent\textsc{Definition} \thethenum. ---  \it  }{\rm }

\newenvironment{remark}{\medbreak\refstepcounter{thenum}\noindent{\it Remark} %
\thethenum. --- }{}

\newenvironment{e-rem}{\medbreak\refstepcounter{thenum}{}%
 \thethenum) }{}
\newenvironment{example}{\medbreak\refstepcounter{thenum}\noindent{\sl
    Example} %
\thethenum. --- }{}

\newenvironment{e-ex}{\medbreak\refstepcounter{thenum}{}%
 \thethenum) }{}
\newenvironment{proof}{\smallbreak\noindent{\sc Proof.} --- \rm}{\quad\bull\smallskip\rm}

\newenvironment{proposition}{\medbreak\refstepcounter{thenum}
\noindent\textsc{Proposition} \thethenum. ---  \it  }{\rm }

\begin{document}

\begin{center}{\LARGE\parindent=0pt{\color{FOR}
Bézout identities and control\\ of the heat equation

}}
\end{center}
\vskip2cm

\hbox to \hsize{\parindent =0pt\hbox to 2.5cm{\hfill}\hss\vbox{\hsize = 7cm {\large François \textsc{Ollivier}}
\bigskip

LIX, UMR CNRS 7161 

École polytechnique 

91128 Palaiseau \textsc{cedex}

France

\smallskip

{\small ollivier@lix.polytechnique.fr}
} \hss}
\vskip 0.3cm

\begin{center}\parindent =0pt 16 Mai 2023
\end{center}
\vfill

{\small

\noindent \textbf{Abstract.} 
  \smallskip
Computing analytic Bézout identities remains a difficult task,
which has many applications in control theory. Flat PDE systems have
cast a new light on this problem. We consider here a simple case of
special interest: a rod of length $a+b$, insulated at both ends and
heated at point $x=a$. The case $a=0$ is classical, the temperature of
the other end $\theta(b,t)$ being then a flat output, with
parametrization $\theta(x,t)=\cosh((b-x)(\partial/\partial
t)^{1/2}\theta(b,t)$.

When $a$ and $b$ are integers, with $a$ odd and $b$ even, the system
is flat and the flat output is obtained from the Bézout identity
$f(x)\cosh(ax)+g(x)\cosh(bx)=1$, the computation of which boils down to
a Bézout identity of Chebyshev polynomials. But this form is not the
most efficient and a smaller expression $f(x)=\sum_{k=1}^{n}
c_{k}\cosh(kx)$ may be computed in linear time.

These results are compared with an approximations by a finite system,
using a classical discretization.

We provide experimental computations, approximating a non rational
value $r$ by a sequence of fractions $b/a$, showing that the power
series for the Bézout relation seems to converge. 

}

\eject

\section*{Introduction}
\addcontentsline{toc}{section}{Introduction} 

\subsection*{Bézout relations and control}
\addcontentsline{toc}{subsection}{Bézout relations and control} 
It is known that the ring of entire functions $\C\langle
z\rangle$ is a unique
factorization domain. Moreover, for any subfield $\K\subset\C$, any
ideal of $\K\langle z\rangle$ that admits a finite basis is a principal
ideal~\cite[th.~9]{Helmer1940}. However explicitly finding Bézout
identities is difficult. This problem is related with many
applicational issues in control theory, such as the design of closed
loop controlers. See \textit{e.g.} Berenstein and
Yger~\cite{Berenstein1989} and the references therein.

This interest became even stronger with the introduction of \emph{flat
systems} in the '90s by Fliess, Lévine, Martin and
Rouchon~\cite{FLMR95,FLMR99,Levine09}. Flat systems are systems the
solutions of which can be parameterized by $m$ arbitrary functions,
called \emph{linearizing outputs},
this parametrization being locally invertible. It is not known if
testing if a non linear system is flat is decidable. This problem is
closely related to \emph{Monge problem}~\cite{Monge1787}, considered by
Cartan~\cite{Cartan1914,Cartan1915} and
Hilbert~\cite{Hilbert1912}. See also Zervos~\cite{Zervos1932}. This
problem is more precisely equivalent to testing \emph{orbital
  flatness}~\cite{FLMR99}, \textit{i.e.} flatness allowing \emph{time
  change}, which amount to transformations that also affect the
independent variable.

Considering finite dimensional systems, one requires that the
parame\-trization only involves a finite number of derivatives, which
also implies the functional unicity of the flat outputs in the single
input case.
One may notice that a notion of
\emph{generalized flatness} has been proposed in the finite
dimensional case, allowing parametrization with an potentially
infinite number of derivations~\cite{Ollivier2022}.

The goal of this paper is to provide a fast algorithm for computing
GCDs of Chebyshev polynomials, that could be used to approximate the
GCD of $\cosh(ax)$ and $\cosh(bx)$ is the general case. This method is
inspired by the control of heat equation or wave equation that suggests
a simple paper tape folding process. One may refer to
\cite{aecf,Gathen1999} for general results on GCD computations. See
Chyzak \textit{et al.}~\cite{Chyzak2005} for computer algebra
algorithms related to parametrizations in differential or Ore
algebras.

\subsection*{Flat control systems}
\addcontentsline{toc}{subsection}{Flat control systems} 
In the ordinary linear case, flatness reduces to controllability,
which means from a mathematical standpoint that the module associated
to the system has no torsion element. In that case, the associated
$\R(t)[\dd/\dd t]$-module is a free module, hence
flat~\cite{FLMR95}. One may notice that, for time varying systems, the flat
parametrization may be undefined where some numerators vanish: flat
systems are generaly understood as admitting a flat parametrization of
a dense open set. See Kaminski \textit{et al.}~\cite{KLO18} for the study of
flat singularities.

Gluing together two flat single
input systems, with linearizing outputs $z_{1}$ and $z_{2}$, we have
$u=L_{1}z_{1}=L_{2}z_{2}$, assuming that the input $u$ is the same for
both systems. This means, in the stationary case where
$L_{i}\in\R[d/d t]$ that we have a parametrization $z_{1}=L_{2}z$
$z_{2}=L_{1}z$ that gives $u=L_{1}L_{2}=L_{2}L_{1}z$, which is
injective and surjective iff $GCD(L_{1},L_{2})=1$. In such a case, the
Bézout identity $M_{1}L_{1}+M_{2}L_{2}=1$ provides the expression of
the new flat output $z=M_{2}z_{1}+M_{1}z_{2}$.

Using Mikusiński's operational calculus~\cite{Mikusinski1959},
flatness has been generalized to linear PDE, considering a robot arm,
with small deflexion, described by the Euler-Bernouilli
equation~\cite{Fliess1997SystmesLS,Aoustin1997}. Here the analogy is
weaker, as the associated module is not free. On reduces to a free
module by enlarging the operator ring with the inverse of some
operator $\pi$, for which there is some freedom of choice. Considering
the heat equation (see Laroche~\cite{LarochePHD,Laroche2000}), there
is a natural choice for a rod heated at one end and insulated at the
other end, for which it is best to chose the temperature of the
insulated end a flat output. One may notice on this example that the
parametrizations are then given by an \emph{entire analytic
  operator}. With such a requirement, for a single input system, the
flat output is unique, up to a multiplication by an entire analytic
operator with entire analytic inverse. A key issue is that fractional
derivatives that appear in intermediate computations disappear at the
end, with a suitable choice of output, so that the parametrization is
also uniquely defined. One may also accept keeping fractionnal
derivatives, used with succes by Oustaloup~\cite{Oustaloup2015}, but
then their definition is not unique, and one may also accept that the
parametrization depends on this choice, as proposed by Rammal
\textit{et al.}~\cite{Rammal2020}.

Flatness has also been generalized to non-linear PDE systems by
reducing to a sequence of finite dimensional flat systems, using
suitable discretizations~\cite{Ollivier2001,Rigatos2015}.

As we see, many theoretical approches are available, and we are far
from a general unified algebraic theory. We will focus here on simple
cases where the use of $\R\langle \dd/\dd t\rangle$-modules is relevant.

\subsection*{Bézout relations for the heat equation}
\addcontentsline{toc}{subsection}{Bézout relations for the heat
  equation}
\label{subsec:Bezout_heat}
We will be concerned here with the problem of computing Bézout
identities for entire analytic operators, and will focus on the case
of $\cosh(at)$ and $\cosh(bt)$, investigating first the case when $b/a$
is rational which reduces to the simple case when $a$ and $b$ have no
common factor. Then, a rod of length $a+b$ insulated at both ends and
heated at point $x=a$ is controllable iff $a$ and $b$ are not both
odd.

The problem then boils down to computing the GCD of Chebyshev
polynomials, which becomes hard for large degrees. However, this may
be done in linear time using a representation of the polynomials in
the\break Bézout identity $M_{1}\cosh(at)+M_{2}\cosh(bt)=1$ as
$$M_{i}(\cosh(t))=\sum_{k}a_{i,k}\cosh(kt)$$ that suits better our
purpose. Then, one may try to consider the general case by using
approximations by discretizations or truncated Fourier series, or by
considering rational approximations of an irrational value of $a/b$,
for which fast computations of Bézout identities for Chebyshev
polynomials with degrees up to $10^{5}$ are usefull.

Our aim is to provide computational tools allowing mathematical
experimentations, resting on Maple implementations.

\subsection*{Plan of the paper}
\addcontentsline{toc}{subsection}{Plan of the paper} 
The plan of the paper is the following. In a section~\ref{sec:flatness_linear}, one
recalls some basic definitions and properties of linear flat systems,
focussing on the single input.

In section~\ref{sec:heat}, we investigate the heat equation for a rod,
first in the case of a rod heated at one end, then reducing the
general case to the case of two rods of different lengths, which leads
to the computation of a Bézout relation for $\cosh$ operators

In a third section~\ref{sec:linear_time}, we describe a linear time algorithm for computing
the Bézout relation, using a suitable representation.

In section~\ref{sec:discr}, we use a discretization of the rod, showing that
its flat output is the same as the flat output coming from the Bézout
relation. Of this we deduce a linear time method, based on folding and cutting
paper tapes.

Experiments of computations are presented in sec.~\ref{sec:discr}

\section{Flatness for linear EDO systems}\label{sec:flatness_linear}

We recall here a few basic results about flat systems in finite
dimension that will be needed in the sequel or help understand the
situation. 

\subsection{General results}\label{subsec:linear_general}

As already stated in the introduction, in the linear case, systems are
best described by $\R(t)[\dd/\dd t]$-modules and then flatness is
equivalent to controllability, using the following classical result,
for which we refer to Jacobson~\cite[chap.~3 th.~18]{Jacobson1943}. We
denote by $[\Sigma]$ the submodule of a module $M$ generated by a
family $\Sigma$.

\begin{theorem}\label{th:Jacobson} If $A$ is a euclidean domain, or
  more generally a principal ideal domain, possibly non commutative,
  then any finitely generated $A$-module $M$ admits a decomposition:
  $M=L\oplus T$, where $L$ is a free module, and $T$ is torsion.
\end{theorem}

Obviously, the existence of a non trivial torsion part means that the
system is non controllable. Reducing to a first order system, as we
may, a torsion element and all its derivatives do no depend on the
inputs, that satisfy no differential equations. So, the system is non
flat. Reciprocally, if $T=0$, the module is free and any basis of $L$
is a flat output, providing a flat parametrization. We denote the
derivation $\dd/\dd t$ by $\dd_{t}$.

\begin{example} Let be the system
$$
  \begin{array}{lll}
    x_{1}'&=&u\\
    x_{2}'&=&u.
  \end{array}
$$
  We associate to it the module $M$, which is the quotient of
  $$\left(\R(t)[\dd_{t}]x_{1}+\R(t)[\dd_{t}]x_{2}+\R(t)[\dd_{t}]u\right)$$
  by its submodule $[\dd_{t}x_{1}-u,\dd_{t}x_{2}-u]$. It is easily
  seen that 
$$
M=[x_{1}]\oplus[x_{1}-x_{2}],
$$
where $[x_{1}]$ is free and $[x_{1}-x_{2}]$ is torsion, as $x_{1}'-x_{2}'=0$.
\end{example}

\begin{example} We now consider the system
$$
  \begin{array}{lll}
    x_{1}'&=&tu\\
    x_{2}'&=&u,
  \end{array}
  $$
  and define $M$ accordingly. It is easily seen as $M$ is now free,
  as $M=[z:=tx_{1}-x_{2}]$. Indeed, $z'=x_{1}$, $z''=u$ and $x_{2}=tz'-z$.  
\end{example}

The following theorem provides a simple criterion for flatness in the
linear case.
\begin{theorem}\label{th:linear_flat} We consider a linear system
  \begin{equation}\label{eq:linear}
    x_{i}'=L_{i}(t,x,u):=\sum_{k=1}^{n}c_{i,k}(t)x_{k}+\sum_{j=1}^{m}d_{i,k}(t)u_{j},\>\hbox{for}\>
    1\le i\le n,
  \end{equation}
where the $x_{i}$ are the state variables and the $u_{j}$ the controls.

We denote by $\partial_{x_{i}}$ the partial derivative
$\partial/\partial x_{i}$, \dots\ The derivation $\dd_{t}$ on the
quotient module is then given by
\begin{equation}\label{eq:dt}
\dd_{t}=\partial_{t}+\sum_{i=1}^{n}L_{i}(t,x,u)\partial_{x_{i}}+
\sum_{j=1}^{m}\sum_{k\in\N}u_{j}^{(k+1)}\partial_{u_{j}^{(k)}}.
\end{equation}
We then define $\Gamma_{0}:=\langle\partial_{u_{1}}, \ldots,
\partial_{u_{m}}\rangle$\footnote{The notation $\langle\Sigma\rangle$
  denote the $\R(t)$-vector space generated by $\Sigma$.}, and then recursively
$\Gamma_{i+1}:=\Gamma_{i}+[\dd_{t},\Gamma_{i}]$.

  With these definitions, the torsion elements are first integrals of
  the derivations in $\Gamma_{n}$, which means that the system is
  flat iff $\dim\Gamma_{n}=n+m$, as a $\R(t)$-vector space.
\end{theorem}
\begin{proof} First we show that torsion elements cannot depend on the
  controls and their derivatives. Indeed, the module is finitely
  generated, so the dimension of the torsion submodule $T$ as a $\R$
  vector space is finite. Assume that $u_{j}^{(k)}$ is the highest
  derivative of $u_{k}$ appearing in the elements of $T$ and that is
  appear in some torsion element $y$. Then $y'$ must depend on
  $u_{j}^{(k+1)}$, using formula~\ref{eq:dt}: a contradiction.

  So torsion elements are first integrals of $\Gamma_{0}$. Now, the
  derivatives of torsion elements are torsion elements and so first
  integrals of $\Gamma_{0}$ too. This means that for $y\in T$,
  $\Gamma_{0}y=0$ and $\Gamma_{0}\dd_{t}y=0$, so that
  $[\dd_{t},\Gamma_{0}]y=0$, so that $\Gamma_{1}y=0$. We can iterate
  the process, showing that $\Gamma_{k}y=0$, for all $k\in\N$. It is
  easily seen that $\dim\Gamma_{k}\le n+m$, so that the sequence
  $\Gamma_{0}\subset\Gamma_{1}\subset\cdots$ must be stationary and
  equal to $\Gamma_{n+m}$. The torsion elements $y$ are then such that
  $\Gamma_{n+m}y=0$.  
\end{proof}

One must stress that in the non linear case, the computation of the
firt integral is much more complicated and that sometimes no rational
or algebraic first integral exists. See Chèze and Combot~\cite{Cheze2020}.

\subsection{The single input case}\label{subsec:single_input}

We give two simple results in the single input case, that we will need
in the sequel.

\begin{theorem}\label{th:Gamma_flat} Let be a linear flat single input
  system~\eqref{eq:linear}, 
  its flat outputs are non trivial first integrals of the
  $\R(t)$-vector space $\Gamma_{n-1}$,
  which is of dimension $n$, so flat outputs linear in the $x_{i}$ are
  unique up to multiplication by a factor.
\end{theorem}
\begin{proof} As the system is flat, $\Gamma_{n}$ must have full
  rank $n+1$, according to th.~\ref{th:linear_flat}. As
  $\dim\Gamma_{i+1}-\dim\gamma_{i}$ is at most the number of controls,
  so $1$, we need have $\dim\Gamma_{i}=i+1$ for all $0\le i\le n$ and
  so $\dim\Gamma_{n-1}=n$. Linear first integrals of $\Gamma_{n-1}$ are
  defined by a linear system of $n$ independent equation in $n+1$
  variables, so that a linear non trivial solution in the $x_{i}$ must exist,
  which is unique, up to multiplication by a constant in $\R$.

  Let $z$ be such a non trivial first integral. The module is free, so
  that $z$, $z'$, \ldots, $z^{n-1}$ are independent. One easily checks
  that $z^{(r)}$ is a first integral of $\Gamma_{n-r-1}$, so that they
  are linear combinations of the $x_{i}$ and first integrals of
  $\Gamma_{0}$, so that we may recompute the $x_{i}$ as linear
  combinations of $z$, \ldots, $z^{(n-1)}$ and then $u$, using
  $z^{(n)}$. This precisely means that $z$ is a flat output.
\end{proof}

Our second results considers gluing two flat systems with the same
single input $u$ and provides an obvious criterion for the
resulting system to be flat. For simplicity, we retreat here to
stationary systems, that is systems with constant coefficients, and the
commutative case, as all systems in the sequel will be of this kind.

\begin{theorem}\label{th:flat_GCD} Let be two flat systems with the
  same control $u$, and flat outputs $z_{1}$ and $z_{2}$. We use the
  derivatives $z_{i}$, \dots, $z_{i}^{(n_{}-1)}$ as state variables
  for system $i=1,2$ and define a module by the two expressions of $u$:
  \begin{equation}\label{eq:gluing}
    u=L_{i}(z_{i}),\>\hbox{for}\> i=1 \>\hbox{or}\> i=2, 
  \end{equation}
  where $L_{i}$ is a linear operator in $\R(t)[\dd_{t}]$ of order $n_{i}$.

  The system \eqref{eq:gluing} is flat iff $\GCD(L_{1},L_{2})=1$.
\end{theorem}
\begin{proof} If $\GCD(L_{1},L_{2})=M$, with $M$ non trivial, then let
  $T_{i}:=L_{i}/M$, for $i=1,2$, and $T:=T_{1}z_{1}-T_{2}z_{2}$. We
  obviously have $MT=0$, so that $T$ is torsion. Reciprocally, if
  $\GCD(L_{1},L_{2})=1$, we have a Bézout relation
  $M_{1}L_{1}+M_{2}L_{2}=1$, so that $z:=M_{1}z_{2}+M_{2}z_{1}$ is a flat
  output for the full system, with a parametrization given by
  $z_{1}=L_{2}z$ and $z_{2}=L_{1}z$.
\end{proof}

\section{The heat equation for a rod}\label{sec:heat}

Considering here a partial differential equation, we need to consider
modules over the ring of entire functions $\R\langle \partial_{t}
\rangle$, or sometimes $\R\langle \partial_{t}^{1/2} \partial\rangle$
during computations.

\subsection{The simple case}\label{subsec:heat_simple}
We consider the heat equation on a rod of length $a$ that is heated at
the end $x=a$, and insulated at $x=0$. We follow the presentation of
Laroche \textit{et al.} This is decribed by the system
\begin{equation}\label{eq:heat_PDE}
  \begin{array}{lll}
    \partial_{t}\theta(x,t) &= &\partial_{x}^{2}\theta(x,t)\\
    \theta(a,t)&=&u(t)\\
    \partial_{x}(0,t) &= &0,
  \end{array}
\end{equation}
denoting $\partial/\partial x$ by $\partial_{x}$, \dots\ In the
Mikusiński domain, one may define $\partial_{t}^{1/2}\theta$, which must
then be equal to $\pm\partial_{x}\theta$. Then, the
general solution is of the form
\begin{equation}\label{eq:gen_sol_heat}
[c_{+}\exp((x-x_{0})\partial_{t}^{1/2})+c_{-}\exp((x-x_{0})\partial_{t}^{1/2})]\theta(x_{0},t),
\end{equation}
with $c_{+}+c_{-}=1$. Choosing $x_{0}=a$, we get
\begin{equation}\label{eq:flat_par_heat}
  \theta(x,t)=\cosh(x\partial_{t}^{1/2})\theta(0,t).
\end{equation}
Indeed, as $\partial_{x}(0,t)=0$,
$\partial_{t}^{p}\partial_{x}\theta(0,t)=\partial_{x}^{2p+1}\theta(0,t)=0$,
so that all odd derivatives vanish at this point in the general
solution~\eqref{eq:gen_sol_heat}. We need choose for the value of the
flat output $\theta(0,t)$ functions $f(t)$ that provide converging
series. It is shown in \cite[th.~1]{Laroche2000} that this is granted
for Gevrey $\alpha$ functions, with $\alpha<2$\footnote{We recall that
  function $f$ Gevrey or order $\alpha$ if there exist $M$ and $R$
  such that for all $m\in\N$ $f^{m}(t)\le M\frac{(m!)^{\alpha}}{R}$.}.

\subsection{The general case}

We consider here the case of the heat equation for a rod of length
$a+b$ insulated at both ends $x=0$ and $x=a+b$, and heated at point
$x=a$, so the control is $u(t)=\theta(a,t)$. We have two copies of
the problem investigated at subsec~\ref{subsec:heat_simple} and we can
rely on the parametrization already found, using
\begin{equation}\label{eq:heat_two}
\begin{array}{lll}
  \theta(x,t)&=&\cosh(x\partial_{t}^{1/2})\theta(0,t)\>\hbox{for}\> 0\le
  x\le a\\
\theta(x,t)&=&\cosh((a+b-x)\partial_{t}^{1/2})\theta(a+b,t)\>\hbox{for}\> a\le
x\le a+b,
\end{array}
\end{equation}
but we need have the compatibility relation
\begin{equation}\label{eq:heat_compat}
  \begin{array}{lll}
    u(t)=\theta(a,t)&=&\cosh(a\partial_{t}^{1/2})\theta(0,t)\\
    &=&\cosh(b\partial_{t}^{1/2})\theta(a+b,t).
   \end{array}
  \end{equation}
We may proceed as done in~\cite{Fliess1997SystmesLS,Aoustin1997}
and allow ourselves to invert some operator. Let
\begin{equation}\label{eq:heat_z}
  z(t)=\acosh(a\partial_{t}^{1/2})^{-1}\theta(a+b,t),
\end{equation}
according to eq.~\eqref{eq:heat_compat}, this implies
\begin{equation}\label{eq:heat_z2}
  z(t)=\cosh(a\partial_{t}^{1/2})^{-1}\theta(a+b,t),
\end{equation}
so that the compatibility condition stands, using the parametrization
\begin{equation}\label{eq:heat_full}
  \begin{array}{lll}
  \theta(x,t)&=&\cosh(x\partial_{t}^{1/2})\cosh(b\partial_{t}^{1/2})\theta(0,t)\>\hbox{for}\> 0\le
  x\le a\\
  \theta(x,t)&=&\cosh((a+b-x)\partial_{t}^{1/2})\cosh(a\partial_{t}^{1/2})\theta(a+b,t)\\
  &&\hbox{for}\> a\le x\le a+b.
\end{array}
\end{equation}
It is easily seen that the operators $\cosh(a\partial_{t}^{1/2})$ and
$\cosh(b\partial_{t}^{1/2})$ have a non trivial GCD iff $\cosh ax$ and
$\cosh bx$ have. This can only happen when $a/b$ is rational. Without
loss of generality, we can reduce with a change of time scale to the
case when $a$ and $b$ are integers without common factors. Then $\cosh
bx=T_{b}(\cosh x)$, so that we are reduced to computing the GCD of
$T_{a}$ and $T_{b}$ which is non trivial iff $a$ and $b$ are odd. In
this case, the function $\hat\theta(x,t)=e^{-\pi^{2} t/4}\cos(\pi
x/2)$ is a solution of the full PDE system and limit conditions, with
$u(t)=\hat\theta(a,t)=0$, so that $\hat\theta$ is torsion:
$\partial_{t}\theta=-\pi^{2}\hat\theta/4$. The PDE system is not
controllable for $a$ and $b$ both odd.

Hence we can focus on the case $a$ even and $b$ odd, for which we have
controllability and can compute a Bézout identity
$L_{1}T_{a}+L_{b}T_{b}=1$ allowing to express the flat output $z$ in the
following way:
\begin{equation}
  \begin{array}{lll}
    z(x,t)=L_{1}(\cosh(x\partial_{t}^{1/2}))\theta(0,t) \>\hbox{for}\> 0\le
  x\le a\\
    z(x,t)=L_{1}(\cosh((a+b-x)\partial_{t}^{1/2}))\theta(a+b,t) \>\hbox{for}\> a\le
  x\le a+b.
  \end{array}
\end{equation}

In fact, we will need to consider accurate rational approximations of
real number $r=b/a$ and so great values of integers $a$ and $b$ for
which a naive computation becomes soon impossible. 

\begin{remark}\label{rem:quadratique}
  To be perfectly rigorous, we work here in the ring $A$ of
  entire differential operators $\R\langle \partial_{t}\rangle$. As
  already stated, any
ideal of $\K\langle z\rangle$ that admits a finite basis is a principal
ideal~\cite[th.~9]{Helmer1940}, so that any finite type $A$-module $M$
admits a decomposition $M=F\oplus T$, where $F$ is free and $T$ is
torsion, according to th.~\ref{th:Jacobson}.
In our case, we consider the quotient
$$
(Ae_{1}+Ae_{2})/[\cosh(b\partial_{t}^{1/2})e_{1}+\cosh(a\partial_{t}^{1/2})e_{2}],
$$ where the generators $e_{1}$ and $e_{2}$ are meant to represent the time
functions $\theta(0,t)$ and $\theta(a+b,t)$, if one wishes to recover
some mathematically non rigorous but easily understood physical interpretation.
\end{remark}

\section{A linear time algorithm}\label{sec:linear_time}

\subsection{Description of the algorithm}\label{subsec:descr_algo}

In the case where $a$ and $b$ are integers such that
$\GCD(T_{a},T_{b})=1$, we are loooking for a Bézout relation
$L_{1}T_{a}(\cosh x)+L_{2}T_{b}(\cosh x)=1$, with $\deg L_{1}\le b-1$
and $\deg L_{2}\le a-1$. We want to use a representation of $L_{1}$
and $L_{2}$ as
\begin{equation}\label{eq:alter_Cheb}
  L_{i}:=\sum_{k=1}^{b-1}c_{i,k}\cosh(kx).
\end{equation}
The basis is to use the classical formula
$2\cosh(ix)\cosh(jx)=\cosh((i+j)x)+\cosh(|i-j|x)$.

\begin{remark}\label{rem:parity}
a) One knows that if $a$ or $b$ is even (resp.~odd), then $T_{a}$ or $T_{b}$
involves only terms of even (resp.~odd) degree so that $L1$ or $L_{2}$
are of even (resp.~odd) degree.

b) As $\deg L_{1}\le b-1$ and $\deg L_{2}\le a-1$, the terms
involved in the Bézout relation are of even degree $k$ with $0\le
k\le a+b-1$.
\end{remark}

\begin{theorem}\label{th:algo}
  Assume that $a$ and $b$ have no common factor and that
  one is odd and the other is even, then there
  exists integer sequences $\alpha_{i}$, $c_{i}$,
  $k_{i}$ and $f_{i}$, for $1\le i\le (a+b+1)/2$, such that for all
  $1\le i_{0}\le (a+b+1)/2$
  \begin{equation}\label{eq:th_algo}
\sum_{i=1}^{i_{0}}c_{i}\cosh(f_{i}x)\cosh(\alpha_{i}x)=1+d_{i_{0}}\cosh(k_{i_{0}}x),
  \end{equation}
  where $\alpha_{i}$ is equal to $a$ or $b$, the $k_{i}$ are even and
  $0\le k_{i}\le a+b-1$, $0\le f_{i}<b$ (resp.~$0\le f_{i}<a$) when
  $\alpha_{i}=a$ (resp.~$b$), $c_{i}=\pm2$ and $d_{i_{0}}=c_{i_{0}}/2$
  if $1\le i_{0}<(a+b+1)/2$, $c_{(a+b+1)/2}=\pm1$ and
  $d_{(a+b+1)/2}=0$, so that the sum is equal to $1$ when
  $i_{0}=(a+b+1)/2$. By convention, we set $k_{0}=0$.
\end{theorem}
\begin{proof} The proof is done by induction on $i_{0}$.
  When $i=1$, the constant term $1$ must come from
  $$2\cosh(ax)\cosh(ax)\>\hbox{or}\>2\cosh(bx)\cos(bx).$$ Assuming that $a<b$,
  as we may up to a permutation, then
  $$
  2\cosh(bx)\cosh(bx)=1+\cosh(2bx)
  $$
  that includes a term of degree $2b>a+b-1$, which is
  excluded. So we need use $2\cosh(ax)\cosh(ax)$ when $a<b$, which
  makes appear a term $\cosh(2ax)$. We set then $f_{1}:=a$, $c_{1}=2$,
  $\alpha_{1}=a$ and $k_{1}=2a$, so that $k_{1}$ is even. (This is
  step 1.\ of algorithm~\ref{algo:Chebyshev_Bezout}.)

Assume that we have \eqref{eq:th_algo} with $k_{i}$, $f_{i}$, $c_{i}$
and $\alpha_{i}$ according to our requirements for all $i$ up to
$i_{0}$.  
  There are $2$ possible values for $0\le k_{i_{0}+1}\le a+b-1$,
  so that
  \begin{equation}\label{eq:requ}
  c_{i_{0}}/2\cosh(k_{i_{0}}-c_{i_{0}}\cosh(f_{i_{0}+1}x)\cosh(\alpha_{i_{0}}
  x)=c_{i_{0}+1}\cosh(k_{i_{0}+1}x).
  \end{equation}

  i) We can always use $k_{i_{0}+1}=|2a-k_{i_{0}}|$, which is such that
  $$c_{i_{0}}/2\cosh(k_{i_{0}}-c_{i_{0}}\cosh(f_{i_{0}+1}x)\cosh(ax)=
  c_{i_{0}+1}\cosh(k_{i_{0}+1}x),$$ 
  providing a contribution of $-c_{i_{0}}\cos(|a-k_{i}|x)$ to
  $L_{1}$. We would set then $\alpha_{i_{0}}=a$,
  $f_{i_{0}+1}:=|a-k_{i_{0}}|$, $c_{i_{0}+1}:=-c_{i_{0}}$ and $d_{i_{0}+1}:=c_{i_{0}+1}/2$.

  ii) a) If $k_{i_{0}}\ge b-a+1$, then we can also choose
  $k_{i_{0}+1}=2b-k_{i}\le b+a-1$, so that
  $$c_{i_{0}}/2\cosh(k_{i_{0}}x)-c_{i_{0}}\cos(|b-k_{i}|x)\cosh(bx)=
  -c_{i_{0}}/2\cosh(k_{i+1}x),$$ providing a contribution of
  $-c_{i_{0}}\cosh(|b-k_{i}|x)$ to $L_{2}$. We could set then
  $\alpha_{i_{0}+1}=b$, $f_{i_{0}+1}:=|b-k_{i}|$,
   $c_{i_{0}+1}:=-c_{i_{0}}$ and $d_{i_{0}+1}:=c_{i_{0}+1}/2$.

  ii) b) If $k_{i_{0}}\le b-a-1$, we
  may set
  $k_{i_{0}+1}=2a+k_{i_{0}}$, so that
  $$c_{i_{0}}/2\cosh(k_{i}x)-c_{i_{0}}\cos((a+k_{i_{0}})x)\cosh(ax)=
  -c_{i_{0}}/2\cosh(k_{i+1}x),$$ 
  providing a contribution of $-c_{i_{0}}\cos((a+k_{i})x)$ to $L_{1}$. We
  could set then $\alpha_{i_{0}+1}=a$, $f_{i_{0}+1}:=a+k_{i_{0}}$,
 $c_{i_{0}+1}:=-c_{i_{0}}$ and $d_{i_{0}+1}:=c_{i_{0}+1}/2$.

  Among these two possible values for $k_{i_{0}+1}$, that are seen to
  be even when $k_{i_{0}}$ is even, one is the value of $k_{i_{0}-1}$,
  so the other value must be chosen for $k_{i_{0}+1}$. We always have
  $k_{i_{0}+1}\neq k_{i_{0}}$ and $k_{i_{0}-1}\neq k_{i_{0}}$, except
  in two cases. The first is $k_{0}=0$, set above by convention, which
  is in fact the value for $k_{2}$ coming from rule i) whith
  $k_{1}=2a$, that the convention $k_{0}=0$ excludes. The second case
  is $k_{i_{0}}=a$ if $a$ is even, in which case i) sets
  $k_{i_{0}+1}=a$ or $b$ if is $b$ is even, in which case ii) a) sets
  $k_{i_{0}+1}=b$. As the sequence $k_{i}$ starts with $k_{0}=0$ that
  is a stationary value, and the only a finite number of values are
  possible for the $k_{i}$, it must end at the second stationary value
  for some $k_{p}=a$ (resp.~$k_{p}=b$) when $a$ (resp.~$b$ is even).

  Then, we only have to set $c_{p}=-c_{i}/2$, so that the sum
  \eqref{eq:th_algo} is equal to $1$ and we may set $d_{p}=0$, the
  choice of $k_{p}$ being then unimportant.

  We only have left to show that the maximal index $p$ is indeed
  equal to $(a+b+1)/2$. Consider the equivalence relation $\equiv$ in $\Z$ such
  that $x\equiv y$ if $x=-y$ or $x-a=-y+a$ or $x-b=-y+b$. According to
  rules i) and ii), for any value $k_{i}$ in the sequence, any value
  $0\le k\le a+b-1$ such that $k\equiv k_{i}$ also belong to the
  sequence. Now, as $\GCD(a,b)=1$ there are only two equivalence
  classes: the class of $0$ and the class of $1$, so that all even
  values $0\le k\le a+b-1$ must belong to the sequence.
  The paper folding process of
  sec.~\ref{subsec:paper} is an illustration of this property.
\end{proof}

From the previous proposition, we deduce the following algorithm,
which we have implemented in a Maple package.

\begin{algorithm}\label{algo:Chebyshev_Bezout}
\textbf{Input} Two integers $a$ and $b$ with $a<b$, one even, the
other odd.

\textbf{Output} The factors $L_{1}$ and $L_{2}$ in a Bézout relation
for the Chebyshev polynomials $T_{a}$ and $T_{b}$, represented by two
arrays $A_{1}$ and $A_{2}$ with $A_{i}[i]=c_{i}$ if $c_{i}\cosh(ix)$
appears in $L_{i}$.

\textbf{Step 1.} $k_{0}:=0$, $k_{1}:=2a$, $A_{1}[a]:=2$;

\textbf{if} $a$ is even then $k_{final}:=a$ \textbf{else}
$k_{final}:=b$ \textbf{fi};  

\textbf{Step 2.} \textbf{while} $k_{i}\neq k_{final}$ \textbf{do}

\hbox to 1cm{\hfill} Determine $\alpha_{i+1}$, $k_{i+1}$, $c_{i+1}$ and $f_{i+1}$ as
in th.~\ref{th:algo} using rules i) or
ii)in the proof.

\hbox to 1cm{\hfill} Set $A_{1,i}[f_{i}]:=c_{i}$ if $\alpha_{i}=a$;

\hbox to 1cm{\hfill} Set $A_{2,i}[f_{i}]:=c_{i}$ if $\alpha_{i}=b$;

\hbox to 1cm{\hfill} $i:=i+1$; \textbf{od};

\textbf{Step 3}. ($k=k_{final}$) \textbf{if}
$a$ is even \textbf{then} $A_{1}[a]:=-c_{(a+b-1)/2}/2$ \textbf{else}
$A_{2}[b]:=-c_{(a+b-1)/2}/2$ \textbf{fi};

\textbf{return} $A_{1}$ and $A_{2}$.
\end{algorithm}

\subsection{Complexity issues and
  implementation}\label{subsec:implementation}

It is easily seen that the total number of operations in
algorithm~\ref{algo:Chebyshev_Bezout} is proportional to the number
$(a+b+1)/2$ of steps, so $O(a+b)$. As the number of steps is also the
number of terms in the output, the complexity is linear in the size of
the result and no great improvement can be expected.

One must notice that a naive use of Maple addition in the previous
algorithm leads to a quadratic complexity, as the cost of addition is
linear, but we just need a power series expansion
up to a chosen order. Computing just the arrays is very fast. 

\begin{figure}[ht!]
  \caption{Left, Maple \texttt{gcdex} function. Right our
    implementation. We compute $\GCD(T_{2i},T_{2i+1})$. Time are given
    in sec, depending on $i$.}
  \hbox to \hsize{\includegraphics[width=6.cm]{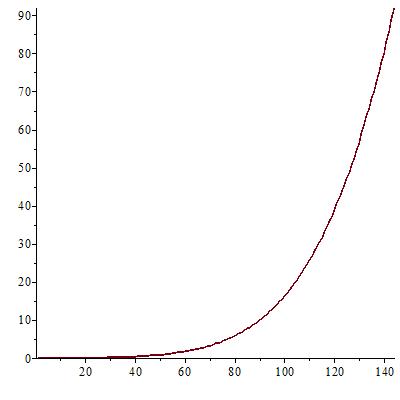}\hfill
    \includegraphics[width=6.cm]{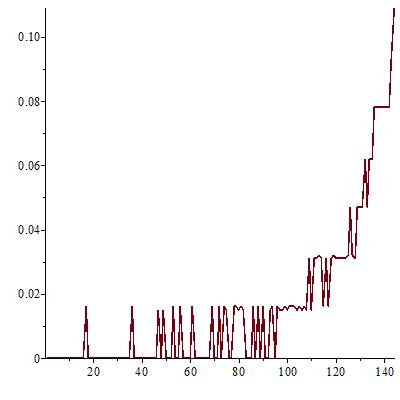}}\label{fig:GCD1}
\end{figure}

We give here a few curves showing CPU time, starting with the
computation of the Bézout relation using Maple \texttt{gcdex} function
and our implementation fig~\ref{fig:GCD1}. We see that our
implementation is much faster for getting the same result. The
irregularities in the right curve is possibly due to the particularity
of Maple's quite unpredictable internal term ordering, implying term
permutations. Of course, just because of the size of intermediate
computations, noticing that the first coefficient of $T_{a}$ is
$2^{a-1}$, general GCD algorithms cannot compete, as they do not use a
suitable data representation. One may notice however that they can
provide already interesting results for pratical purpose in an
acceptable time. 

The following curve on the right exhibits the quadratic behaviour obtained by
computing explicitly the factors $L_{i}$ of the Bézout relation. On
the left, we compute the power series development of the factors, up
to order $20$, and the complexity keeps linear.

\begin{figure}[ht!]
  \caption{Left, we compute the factors $L_{1}$, leading to a
    quadratic complexity. Right, we only compute their power series
    development, up to order $20$. The example is again
    $\GCD(T_{2i},T_{2i+1})$. Time in sec, depending on $i$.}  \hbox to
  \hsize{\includegraphics[width=6.cm]{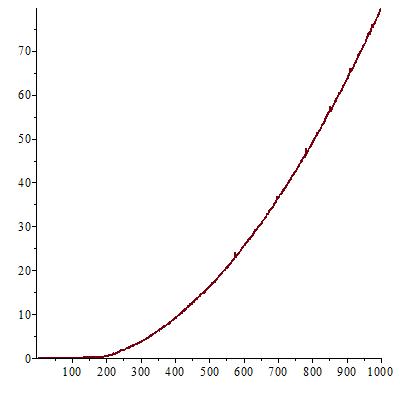}\hfill
    \includegraphics[width=6.cm]{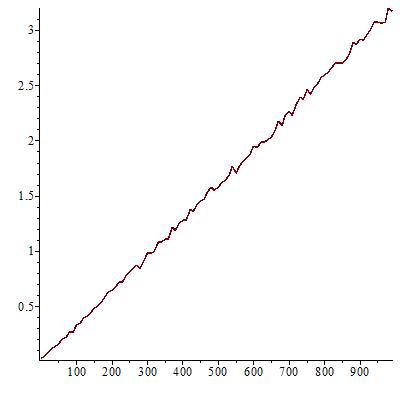}}\label{fig:GCD2}
\end{figure}

The algorithm~\ref{algo:Chebyshev_Bezout} has been implemented in a
function \texttt{BezoutBis} of a Maple package \texttt{Chaleur},
with a few related functions. Global variables \texttt{use\_pol} and
\texttt{use\_ser}, set to \textit{true} or \textit{false} allow to
compute or not the result as a sum of $\cosh(ix)$ or as the
corresponding series. With this implementation, we can reach degrees
up to $10^{6}$, just computing the arrays of coefficients or up to
$10^{5}$ computing power series of order $20$ in a few minutes. 

The Maple package is available at
adress:\hfill\break
\url{http://www.lix.polytechnique.fr/~ollivier/Chaleur/}

\section{Discretization}\label{sec:discr}

Assuming that $a$ and $b$ have no common factor, we use  a classical
parametrization, that is adapted from~\cite{Ollivier2001}:
\begin{equation}\label{eq:discr}
\theta_{i}'=2\theta_{i}-\theta_{i-1}-\theta_{i+1},\>\hbox{for}\>
0\le i\>qa\> \hbox{and}\> qa<i\le q(a+b),
\end{equation}
 with $\theta_{qa}=u$ and, by convention,
$\theta_{-1}=\theta_{1}$ and $\theta_{q(a+b)+1}=\theta_{q(a+b)-1}$.

We will show that this discretized model is flat when $a$ and $b$ are
not both odd and that its flat output
corresponds to the one obtained in the preceeding section for the PDE
system~\ref{eq:heat_PDE}.

For this, one may consider the following system:
\begin{equation}\label{eq:Che}
\begin{array}{lll}
  \theta_{i}'&=&\theta_{i+1},\>\hbox{for}\>
i=0,\\
  \theta_{i}'&=&\theta_{i-1},\>\hbox{for}\>
i=a+b,\\
  \theta_{i}'&=&(\theta_{i-1}+\theta_{i+1})/2,\>\hbox{for}\>
  0\le i\>qa\> \hbox{and}\> qa<i\le q(a+b).
  \end{array}
\end{equation}

\begin{remark}\label{rem:Che}
It is build so that $\theta_{1}=\dd_{t}\theta_{0}$ and
$\theta_{n+1}=2\dd_{t}\theta_{n}-\theta_{n-1}$, where we recognize the
classical recurrence defining Chebyshev polynomials. So we have
$\theta_{i}=T_{i}(\dd_{t})\theta_{0}$, for $0\le i\le a$ and in the
same way $\theta_{a+b-i}=T_{i}(\dd_{t})\theta_{a+b}$.
\end{remark}

Then, easy computations show that for both systems the sets
$\Gamma_{i}$ are the same. More precisely, we have the following
proposition.
\begin{proposition}\label{prop:gamma_i}
  For systems \eqref{eq:Che} and \eqref{eq:discr}, assuming $a<b$, the
  sets $\Gamma_{i}$ are such that
  $\Gamma_{0}:=\langle\partial_{u}\rangle$ and $\Gamma_{i}=\langle c_{1,i}\partial_{x_{k_{1,i}}}+\partial_{x_{k_{2,i}}}$,
  for $1\le i\le a+b$ , where $k_{2,i}=a+i$ for $1\le i\le b$ and 
$k_{2,i}=a+2b-i$, for $b\le i\le a+b$. For $c_{1,i}$ and $k_{1,i}$ the
  rule is the following: $c_{1,i}=0$ for $a=2pa$. Assume that
  $i=2pa+k$ for $p\in\N$ and $<\le k<2a$. If $p$ is odd, then
  $c_{1,i}=1$, if not $c_{1,i}=-1$. If $1<k\le a$, then $k_{1,i}=a-k$
  and if $a\ge k<2a$, then $k_{1,i}=k-a$.
\end{proposition}

\begin{theorem} Let $L_{1}T_{qa}+L_{2}T_{qb}=1$ be a Bézout relation
  for the Chebyshev polynomials $T_{a}$ and $T_{b}$, $a$ and $b$
  without common factor and not
  both odd, with
  $L_{i}=\sum_{k=0}^{b-1}c_{i,k}\cosh(kqx)$, then a flat output for the
  discrete system~\eqref{eq:discr} is
  \begin{equation}\label{eq:flat_output_discr}
    \sum_{k=0}^{a-1}c_{2,k}x_{q(a+b-k)}+\sum_{k=0}^{b-1}c_{1,k}x_{qk}.
  \end{equation}
\end{theorem}
\begin{proof}
By prop.~\ref{prop:gamma_i}, the systems \eqref{eq:Che} and
\eqref{eq:discr} have the same sets $\Gamma_{i}$, and so, according to
th.~\ref{th:Gamma_flat}, the same flat outputs. By rem.~\ref{rem:Che}
and prop.~\ref{th:flat_GCD}, a flat output of \eqref{eq:Che} is \eqref{eq:flat_output_discr}.
\end{proof}

\subsection{Analogy with the wave equation}

One may view the propagation of the indices as a wave, starting at the
heated point, that reflects
on the insulated end. When it goes back to the heated point, then it
reflects too, but with an opposite sign.

This may be easier to understand using an
analogy with the wave equation, which has the same flat output.

\begin{equation}\label{eq:wave_PDE}
  \begin{array}{lll}
    \partial_{t}\theta(x,t) &= &\partial_{x}^{2}\theta(x,t)\\
    \theta(a,t)&=&u(t)\\
    \partial_{x}(0,t) &= &0,
  \end{array}
\end{equation}

Such a system is a delay system, with a flat
parametrization, meaning that the associated module is free:
$\theta(x,t)=\cosh(x\dd_{t})\theta(0,t)=\theta(0,t-x)+\theta(0,t+x)$. Indeed,
in the theory of Mikusiński~\cite{Mikusinski1959}, the operator
$\exp(\dd_{t})$ is a delay operator and
$\exp(x\dd_{t})f(t)=f(t+x)$. See Mounier \textit{et
  al.}~\cite{Mounier1998,Woittenek2010} for more detail on wave control.

As explained in rem.~\ref{rem:quadratique}, we work on the ring $A$
of integer differential operators $\R\langle \partial_{t}\rangle$ and
consider the quotient module
$$
M:=Ae_{1}+Ae_{2}/[\cosh(b\partial_{t})e_{1}+\cosh(a\partial_{t})e_{2}],
$$ where the generators $e_{1}$ and $e_{2}$ are meant again to
represent the time functions $\theta(0,t)$ and $\theta(a+b,t)$. In
this more rigorous setting, the module $M$ is indeed free when $a/b$
is not the quotient of two odd integers. 

\begin{remark} A naive discretization, such as:
\begin{equation}\label{eq:wave_discr}
\theta_{i}''=2\theta_{i}-\theta_{i-1}-\theta_{i+1},\>\hbox{for}\>
0\le i\>qa\> \hbox{and}\> qa<i\le q(a+b),
\end{equation}
would be flat, allowing to reach any point in state space in any non
zero time. So it fails to model the incompressible delay for wave
propagation. See Zuazua~\cite{Zuazua2005} for such issues.
\end{remark}

\subsection{Computations with a paper tape}\label{subsec:paper}

This section may sound anachronical, but as designing new physical
devices for computations is not devoid of interest, a short
presentation of a this simple computational tool may help to
understand the basic idea of the algorithm and as a contribution to
the study of computing as a physical process, even if we do not
actually want to use it!

\begin{figure}[ht!]
  \caption{A tape of paper with boxes, indices on both side of
    borders, one face $+$ and one face $-$}
  \hbox to \hsize{\includegraphics[width=6.cm]{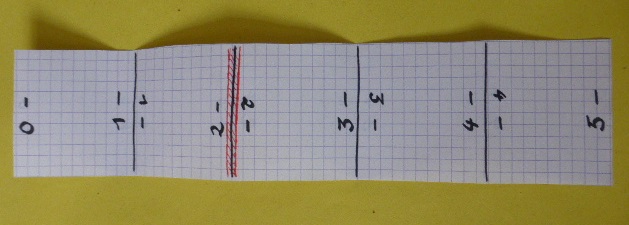}\hfill
    \includegraphics[width=6.cm]{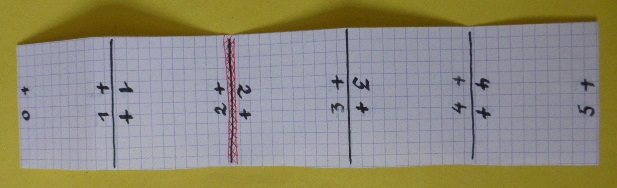}}\label{fig:folding1}
\end{figure}

The process, based on the propagation of the differential operators in
prop.~\ref{prop:gamma_i}, is indeed close to the wave equation that is
equivalent to algorithm~\ref{algo:Chebyshev_Bezout} and may 
help to visualize how the computation of a flat
output using a paper tape divided in $a+b$
boxes. On each side of the border between boxes, we write the index
$i$ and a sign, which is always $+$ on one side and $-$ on the other
side, as shown in fig.~\ref{fig:folding1}. We show here the
computation of the GCD of $T_{2}$ and $T_{3}$ that gives: 
\begin{equation}\label{eq:folding}
(2\cosh(2x)+1)T_{2}(x)-2\cosh(x)T_{3}(x).
\end{equation}

\begin{figure}[ht!]
  \caption{One folds the tape at the heated point in red and cuts the
    part that oversets the shorter end.}
  \hbox to \hsize{\includegraphics[width=6.cm]{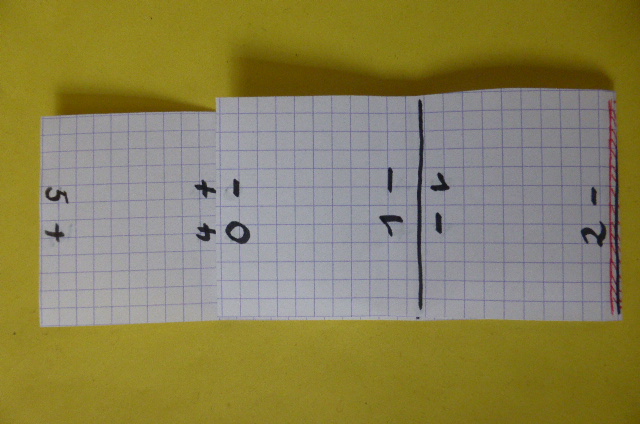}\hfill
    \includegraphics[width=6.cm]{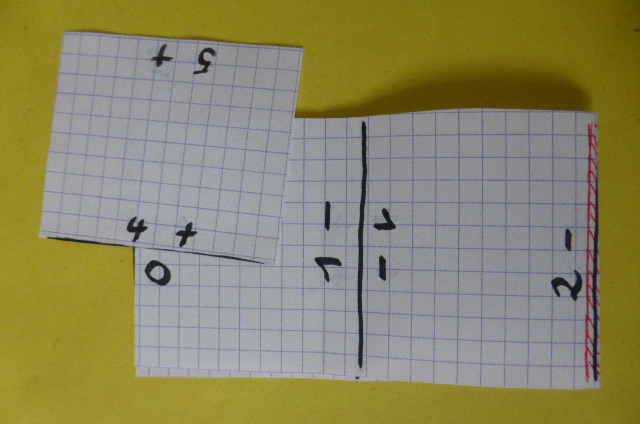}}\label{fig:folding2}
\end{figure}

We mark the heated point with a red line, and fold the paper tape at
this place. During the process, we have a long end and a short end. If
the long end oversets the heated point, we fold it. If it oversets an
end, we cut it and rotate it of $\pi$~rad in the same plane, as shown
in fig.~\ref{fig:folding2}.

We iterate the process until both ends have the same length, which is
the GCD of $a$ and $b$. Then, we look the end opposite to the heated
point, so the odd end, if $a$ and $b$ have no common factor. The
number of $+$ or $-$ for the written indices provide the requested coefficients
of the flat output, up to the sign. The values on the picture show the
opposite of~\ref{eq:folding}. 

\begin{figure}[ht!]
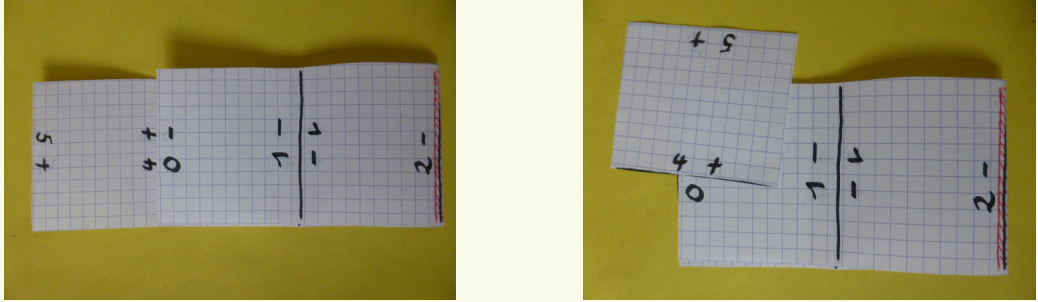

  \caption{We rotate the cut part of $\pi$ and repeat the process. The
    sum of signs at the odd end give the result.}
  \hbox to \hsize{\includegraphics[width=6.cm]{ph3.jpg}\hfill
    \includegraphics[width=6.cm]{ph4.jpg}}\label{fig:folding3}
\end{figure}

Folding corresponds to the change of sign in reflection passing by
index $a$ in prop.~\ref{prop:gamma_i}.

\section{Computational investigations}\label{sec:computations}

Numerical simulations are used to provide empirical estimations of the
Bézout relations for operators $\cosh(x)$ and $\cosh(rx)$, with $r$
irrational, using rational approximations of $r$.  Our running example
is $r=\sqrt{2}$.

The rational approximations used are provided by the continued
fraction expansion:
$$\sqrt{2}=1+\cfrac{1}{2+\cfrac{1}{2+\cfrac{1}{2+\dots}}}$$
using Maple implementation in the package \texttt{NumberTheory}. Of
them, we extract fractions $b/a$ where $b$ and $a$ are not both odd,
so that $\GCD(T_{a},T_{b})=1$. We considered values in this list:
$$
\frac{3}{2},\quad \frac{17}{12}, \frac{99}{70},\quad \frac{577}{408}, \frac{3363}{2378}, \frac{19601}{13860}, \frac{114243}{80782}.
$$ and compute the power series corresponding to the factors in the
Bézout relation, using $20$ digits floats, after substituting $x/a$ to
$x$. We get:
\begin{equation}\label{eq:Bezout_series_1}
  \begin{array}{clll}
    \frac{3}{2}& L_{1}=3.&+1*x^2&+.83x^{4}+\cdots\\
    & L_{2}=-2.&-.25*x^2&-.52*10^{-2}*x^4+\cdots\\
    \frac{17}{12}&L_{1}=1.&-1.33*x^2&-.28*x^4+\cdots\\
    &L_{2}=&.83*x^2&+.72*10^{-1}*x^4+\cdots
  \end{array}
\end{equation}
The two results already look very different, as we are expecting a
convergent sequence. This is due to the fact that $a=2$ is even, but not
a multiple of $4$, so that the constant term in $T_{2}$ is $-1$, while  in
$T_{12}$, which is a multiple of $4$, the constant term is
$1$. We have obtained two seemingly converging sequences, according to the case.
Considering polynomials, the GCDs are normalized by imposing
bounds on the degree, which does not work considering GCDs of integer
analytic functions. The obtained results can only be interpreted as
``convergent'' modulo the trivial relation
$$\cosh(\sqrt{2}x)\cosh(x)-\cosh(x)\cosh(\sqrt{2}x)=0.$$
We have chosen to normalize the relations, so that $L_{2}$ has a
constant term equal to $0$, which is the case for $a$ mutiple of $4$.

In this way, we have results that seem to converge and give the
following estimations for the series, using the approximation
$\frac{114243}{80782}$:
{
$$\begin{array}{lll}
L_{1}&=
&1.-1.3333333332822534857*x^2\\&&-.27941176475495906369*x^4\\
&&-.17992011626456865856e-1*x^6\\&&-.56399934990584172717*10^{-3}*x^8\\
&&-.10660593920209884517*10^{-4}*x^{10}\\&&-.13660155439036212056*10^{-6}*x^{12}\\
&&-.12773250009420268932*10^{-8}*x^{14}\\&&-.9172810105402920608*10^{-11}*x^{16}\\
&&-.5253788359821529842*10^{-13}*x^{18}\\&&-.24709351400071579999*10^{-15}*x^{20},
\end{array}
$$
$$\begin{array}{lll}
L_{2}&=
&.83333333328225349697*x^2\\&&+.71078431383315849009*10^{-1}*x^4\\
&&+.18972403780540266659*10^{-2}*x^6\\&&+.26307421016151944601*10^{-4}*x^8\\
&&+.23019494752812483754*10^{-6}*x^{10}\\&&+.14160103021824419386*10^{-8}*x^{12}\\
&&+.6571055231725949143*10^{-11}*x^{14}\\&&+.24170000965421789288*10^{-13}*x^{16}\\
&&+.7295669958800853034*10^{-16}*x^{18}\\&&+.18497349717353377031*10^{-18}*x^{20}.
\end{array}
$$
}
The total real computation time for this fraction is $428.17$~sec.

Further computations would be needed to investigate the convergence of
these series. The terms are decreasing, but not so fast.

The following table~\ref{fig:table} provides the last term of $L_{1}$,
corresponding to degree $20$ for all our approximations

\begin{figure}[ht!]
\caption{Terms of degree $20$ in $L_{1}$.}
$$
\begin{array}{|c|l|}
  \hline
  {3}/{2}& .27327502044120918666 {\rm E}-15\\
  \hline
  {17}/{12}& .27632394776909554494 {\rm E}-15\\
  \hline
  {99}/{70}& .24164130261737767018 {\rm E}-15\\
  \hline
  {577}/{408} & .24708886190685028010 {\rm E}-15\\
  \hline
  3363/2378 & .24708886190685028010 {\rm E}-15\\
  \hline
  19601/13860 & .24709351973896486850 {\rm E}-15\\
  \hline
  114243/80782 & .24709351400071579999 {\rm E}-15\\
  \hline
\end{array}
$$
\label{fig:table}
\end{figure}

\section*{Conclusion}
\addcontentsline{toc}{section}{Conclusion} 
At this stage, we have been able to design an empirical process to
look for power series expansions of a very specific class of entire
analytic functions, relying on remarkable identities. Such
computational tools exceed the needs of practical control theory but may cast
some light on some theoretical control problem.

On the other hand, control theory provides intuitions to address the
problem of Bézout relations on a wider setting, trying to work in a
direct way with discretizations or Fourier series expansions, topics
on which we already started some investigations, not conclusive at
this stage. Without the comfort of theoretical methods to check of the
validity of computations, being able to get compatible results by
independent ways is essential for experimentations that may help to
improve and make more robust the definitions of controlability and
flatness available for PDE systems.

One thing is clear: data representation is essential for
computational complexity.

\bibliographystyle{amsplain}
\addcontentsline{toc}{section}{References} 
\bibliography{Ollivier-ISSAC.bib}

\providecommand{\bysame}{\leavevmode\hbox to3em{\hrulefill}\thinspace}
\providecommand{\MR}{\relax\ifhmode\unskip\space\fi MR }
\providecommand{\MRhref}[2]{%
  \href{http://www.ams.org/mathscinet-getitem?mr=#1}{#2}
}
\providecommand{\href}[2]{#2}
\begin{thebibliography}{10}

\bibitem{Aoustin1997}
Y.~Aoustin, M.~Fliess, H.~Mounier, P.~Rouchon, and J.~Rudolph, \emph{Theory and
  practice in the motion planning and control of a flexible robot arm using
  mikusinski operators}, Fifth IFAC Symposium on Robot Control (Nantes,
  France), 1997, pp.~287--293.

\bibitem{Berenstein1989}
Carlos~Alberto Berenstein and Alain Yger, \emph{Analytic bezout identities},
  Advances in Applied Mathematics \textbf{10} (1989), 51--74.

\bibitem{aecf}
Alin Bostan, Frédéric Chyzak, Marc Giusti, Romain Lebreton, Grégoire Lecerf,
  Bruno Salvy, and Éric Schost, \emph{Algorithmes efficaces en calcul formel},
  Frédéric Chyzak (auto-edit.), Palaiseau, September 2017 (french), 686
  pages. Printed by CreateSpace.

\bibitem{Cartan1914}
Élie Cartan, \emph{Sur l'équivalence absolue de certains systèmes
  d'équations différentielles et sur certaines familles de courbes}, Bulletin
  de la Société Mathématique de France \textbf{42} (1914), 12--48.

\bibitem{Cartan1915}
\bysame, \emph{Sur l'intĂégration de certains systèmes indéterminés
  d'équations diffĂ©rentielles}, Journal fűr die reine und angewandte
  Mathematik \textbf{145} (1915), 86--151.

\bibitem{FLMR95}
M.~Fliess, J.~L\'{e}vine, Ph. Martin, and P.~Rouchon, \emph{Flatness and defect
  of non-linear systems: introduction theory and examples}, Int. Journal of
  Control \textbf{61} (1995), no.~6, 1327--1361.

\bibitem{FLMR99}
\bysame, \emph{A {L}ie-{B}\"{a}cklund approach to equivalence and flatness of
  nonlinear systems}, IEEE Trans. Automatic Control \textbf{44} (1999), no.~5,
  922--937.

\bibitem{Fliess1997SystmesLS}
Michel Fliess, Hugues Mounier, Pierre Rouchon, and Joachim Rudolph,
  \emph{Syst{\`e}mes lin{\'e}aires sur les op{\'e}rateurs de mikusinski et
  commande d'une poutre flexible}, Esaim: Proceedings \textbf{2} (1997),
  183--193.

\bibitem{Chyzak2005}
Chyzak Frédéric, Quadrat Alban, and Robertz Daniel, \emph{Effective
  algorithms for parametrizing linear control systems over ore algebras}, AAECC
  \textbf{16} (2005), 319--376.

\bibitem{Cheze2020}
Chèze Guillaume and Combot Thierry, \emph{Symbolic computations of first
  integrals for polynomial vector fields}, Found Comput Math \textbf{20}
  (2020), 681--752.

\bibitem{Helmer1940}
Olaf Helmer, \emph{Divisibility properties of integral functions}, Duke
  Mathematical Journal \textbf{6} (1940), 345--356.

\bibitem{Hilbert1912}
David Hilbert, \emph{Ăśber den begriff der klasse von
  differentialgleichungen}, Math.~Annalen \textbf{73} (1912), 95--108.

\bibitem{Jacobson1943}
Nathan Jacobson, \emph{The theory of rings}, Mathematical Surveys and
  Monographs 002, American Mathematical Society, 1943.

\bibitem{KLO18}
Y.~Kaminski, J.~L\'evine, and F.~Ollivier, \emph{Intrinsic and apparent
  singularities in differentially flat systems, and application to global
  motion planning}, Systems \& Control Letters \textbf{113} (2018), 117--124.

\bibitem{LarochePHD}
B.~Laroche, \emph{Extension de la notion de platitude à des systèmes décrits
  par des équations aux dérivées partielles linéaires}, Ph.D. thesis, Ecole
  Nationale Sup\'{e}rieure des Mines de Paris, Paris, France, 2000.

\bibitem{Laroche2000}
B.~Laroche, Ph. Martin, and P.~Rouchon, \emph{Motion planning for the heat
  equation}, Int. J. Robust Nonlinear Control \textbf{10} (2000), no.~8,
  629–643.

\bibitem{Levine09}
J.~L\'{e}vine, \emph{Analysis and control of nonlinear systems: A
  flatness-based approach}, Mathematical Engineering, Springer, Dordrecht,
  Heidelberg, London, New-York, 2009.

\bibitem{Mikusinski1959}
Jan Mikusiński, \emph{Operational calculus}, Pergamon Press, London-New
  York-Paris-Los Angeles, 1959.

\bibitem{Monge1787}
Gaspard Monge, \emph{Supplément où l'on fait savoir que les équations aux
  différences ordinaires, pour lesquelles les conditions d'intégrabilité ne
  sont pas satisfaites sont susceptibles d'une véritable intégration et que
  c'est de cette intégration que dépend celle des équations aux différences
  partielles élevées.}, Histoire de l'Académie royale des sciences, année
  MDCCLXXXIV (1787), 502--576.

\bibitem{Mounier1998}
H.~Mounier, J.~Rudolph, M.~Fliess, and P.~Rouchon, \emph{Tracking control of a
  vibrating string with an interior mass viewed as delay system}, ESAIM:
  Control, Optimisation and Calculus of Variations \textbf{3} (1998), 315--321
  (en). \MR{1644431}

\bibitem{Ollivier2022}
Fran{\c c}ois Ollivier, \emph{{Extending Flat Motion Planning to Non-flat
  Systems. Experiments on Aircraft Models Using Maple}}, {International
  Symposium on Symbolic and Algebraic Computation (ISSAC)} (Lille, France), ACM
  Press, July 2022, pp.~499--507.

\bibitem{Ollivier2001}
François Ollivier and Alexandre Sedoglavic, \emph{A generalization of flatness
  to nonlinear systems of partial differential equations. application to the
  command of a flexible rod}, IFAC Proceedings Volumes \textbf{34} (2001),
  no.~6, 219--223, 5th IFAC Symposium on Nonlinear Control Systems 2001, St
  Petersburg, Russia, 4-6 July 2001.

\bibitem{Rammal2020}
Rim Rammal, Tudor-Bogdan Airimitoaie, Pierre Melchior, and Franck Cazaurang,
  \emph{Unimodular completion for computation of fractionally flat outputs for
  linear fractionally flat systems}, 21st IFAC World Congress (Berlin,
  Germany), IFAC PapersOnLine, vol.~53, IFAC, Elsevier, July 2020,
  pp.~4415--4420.

\bibitem{Rigatos2015}
Gerasimos Rigatos and Alexey Melkikh, \emph{Boundary control of nonlinear pde
  dynamics with the use of differential flatness theory}, 2015 19th
  International Conference on System Theory, Control and Computing (ICSTCC),
  2015, pp.~384--389.

\bibitem{Oustaloup2015}
Jocelyn Sabatier, Patrick Lanusse, Pierre Melchior, and Alain Oustaloup,
  \emph{Fractional order differentiation and robust control design: Crone,
  h-infinity and motion control}, Intelligent Systems, Control and Automation:
  Science and Engineering, vol.~77, springer, Netherlands, 2015.

\bibitem{Gathen1999}
Joachim von~zur Gathen and J\"urgen Gerhard, \emph{{M}odern {C}omputer
  {A}lgebra}, Cambridge University Press, 1999.

\bibitem{Woittenek2010}
Frank Woittennek and Hugues Mounier, \emph{Controllability of networks of
  spatially one-dimensional second order pdes---an algebraic approach}, SIAM
  Journal on Control and Optimization \textbf{48} (2010), no.~6, 3882--3902.

\bibitem{Zervos1932}
Panajiotis Zervos, \emph{Le problème de monge}, Gauthier-Villars, 1932 (fre).

\bibitem{Zuazua2005}
Enrique Zuazua, \emph{Propagation, observation, and control of waves
  approximated by finite difference methods}, SIAM Review \textbf{47} (2005),
  no.~2, 197--243.

\end{thebibliography}

\end{document}